\RequirePackage{fix-cm}
\documentclass[twocolumn]{svjour3}          
\smartqed  
\usepackage{graphicx}
\usepackage{epstopdf} 
\usepackage{epsfig}
\usepackage{amsmath}
\usepackage{bbding}
\usepackage{color}

\begin{document}

\title{Measuring the transition between nonhyperbolic and hyperbolic regimes in open Hamiltonian systems}

\author{Alexandre R. Nieto   \and Euaggelos E. Zotos \and 
     Jes\'{u}s M. Seoane   \and Miguel A. F. Sanju\'{a}n\footnotemark[1] \footnotemark[2]
}


\institute{A. R. Nieto \and J. M. Seoane \and M. A. F. Sanju\'{a}n \at Nonlinear Dynamics, Chaos and Complex Systems Group, Departamento de
              F\'{i}sica, Universidad Rey Juan Carlos, Tulip\'{a}n s/n, 28933 M\'{o}stoles, Madrid, Spain \\
              \email{alexandre.rodriguez@urjc.es} \\
           \and
           M. A. F. Sanju\'{a}n \at Department of Applied Informatics, Kaunas University of Technology, Studentu 50-415, Kaunas LT-51368, Lithuania
           \and
           E. E. Zotos \at
              Departament of Physics, School of Science, Aristotle University of Thessaloniki, GR-541 24, Thessaloniki, Greece
}

\date{Received: date / Accepted: date}

\maketitle
\footnotetext [1]{Corresponding Author : miguel.sanjuan@urjc.es}
\footnotetext [2]{ORCID: 0000-0003-3515-0837}

\begin{abstract}
We show that the presence of KAM islands in nonhyperbolic chaotic scattering has deep implications on the unpredictability of open Hamiltonian systems. When the energy of the system increases the particles escape faster. For this reason the boundary of the exit basins becomes thinner and less fractal. Hence, we could expect a monotonous decrease in the unpredictability as well as in the fractal dimension. However, within the nonhyperbolic regime, fluctuations in the basin entropy have been uncovered. The reason is that when increasing the energy, both the size and geometry of the KAM islands undergo abrupt changes. These fluctuations do not appear within the hyperbolic regime. Hence, the fluctuations in the basin entropy allow us to ascertain the hyperbolic or nonhyperbolic nature of a system.
In this manuscript we have used continuous and discrete open Hamiltonian systems in order to show the relevant role of the KAM islands on the unpredictability of the exit basins, and the utility of the basin entropy to analyze this kind of systems.  

\end{abstract}

\section{Introduction} 
\label{sec:Introduction}
Sensitive dependence on initial conditions is one of the hallmarks of chaos, and is responsible for the unpredictability inherent to the chaotic systems. However, unpredictability has many facets, and for each of them several tools and methods have been developed. One of the facets of unpredictability is the dificulty to predict the evolution of the trajectories. With this perspective, several measures have been developed, such as the topological entropy \cite{Adler}, the expansion entropy \cite{Hunt} and the Kolmogorov-Sinai entropy \cite{Kolmogorov,Sinai}. However, in many physical situations we are interested in the asymptotic behavior rather than the evolution of the system. In this case we consider another facet of unpredictability: the difficulty to predict the final state of a system from certain initial conditions. Under this consideration, basins of attraction \cite{NusseD,OttD} and exit basins \cite{Contopoulos02} have aroused much the interest about the predictability of dynamical systems. A basin of attraction of a dissipative system is the set of initial conditions that are attracted to a certain atractor. Similarly, we define the exit basins in conservative systems as the set of initial conditions that after a finite time escape through one the exits of the system (openings in the potential in continuous systems or predefined regions in area-preserving maps). When two different attractors (or exits) coexist in the phase space, two basins exists and are separated by a boundary. This boundary between the basins can be a smooth curve, but also a fractal curve with non-integer dimension.\\\indent
 In real systems such as engineering systems, the destination of some initial condition is not the unique concern, because the environment is not free of noise and imperfections. Even if in absence of perturbations the basin of attraction can exhibit compact and extensive safe regions, a fast erosion of the basin can occur under small changes in the parameters \cite{Soliman90,Soliman92}. In this way some works developed measures of the dynamical integrity \cite{Thompson89} of the basins of attraction, in order to quantify the capability of the system when accommodating small perturbations without undesired effects \cite{Rega15}. Among other measures, we highlight anisometric local integrity measure (ALIM) \cite{Belardinelli18}, the integrity factor (IF) \cite{Rega05} and the local integrity measure (LIM) \cite{Soliman89}   \\\indent
In this work we study the KAM islands in conservative systems, so we are not interested in the evolution of the trajectories nor in the effect of small perturbations. The escape dynamics of an open Hamiltonian system will vary with the energy (or another parameter of interest). Hence we can analyze the changes on the escape dynamics by simply studying the exit basins for different values of the energy. If we are interested in the fractality of the boundaries of the exit basins, we can calculate the fractal dimension using the uncertainty algorithm \cite{Grebogi83,McDonald85}. However, in order to quantify the unpredictability in this kind of problems we must give an account of the main sources of unpredictability in the exit basins: the number of destinations, the boundary size, and its fractality. None of these factors imply by themselves a high unpredictability. For example, we can deal with a system with a really fractal but thin basin boundary. This system is highly predictable. In order to obtain a new quantitative measure of the unpredictability of the exit basins (or basins of attraction in dissipative systems), recently the basin entropy \cite{Daza16} has been introduced. The basin entropy gives an account of the three previously mentioned ingredients and allows the comparison of the unpredictability of two or more basins. This tool has been used in problems concerning relativistic \cite{Bernal18} and classical chaotic scattering \cite{Nieto18}, and experiments with cold atoms \cite{Daza_cold}. For simplicity, from now on we will use the term unpredictability in the basin entropy sense.

\indent If the escape dynamics of the system is hyperbolic, when the energy increases, the exits widen and the particles escape faster, following an exponential decay law of the survival probability. Therefore, the boundaries of the exit basins become thinner and less fractal. However, if the escape dynamics is nonhyperbolic, the decay law of the survival probability is algebraic and, moreover, there will be trajectories that never escape, following a quasiperiodic orbit that belongs to a Kolmogorov-Arnold-Moser (KAM) torus \cite{Ott}. The quasiperiodic orbits constitute a new destination of the dynamical system and, therefore, appear in the exit basins forming what we know as KAM islands. Analogously to the exit basins, a KAM island is the set of initial conditions that leads to trajectories that do not escape from the scattering region. One of the main motivations for this work is to clarify the effect of the KAM islands on unpredictability of the exit basins. For this purpose, we have selected three open Hamiltonian systems, two continuous and one area-preserving map with escapes, and we have quantified the unpredictability of the exit basins as a function of the energy (or another relevant parameter of the system) using the basin entropy. We have also obtained the fractal dimension to establish whether both quantities provide the same information.
\\\indent
In hyperbolic cases both the fractal dimension and the basin entropy evolve monotonously. However, in the nonhyperbolic case, large fluctuations in the basin entropy appear due to the metamorphosis of KAM islands.
\\\indent The structure of this manuscript is as follows. In Sec.~\ref{secUM} we explain in detail the theoretical and computational aspects of the unpredictability measures that we have used, in particular the basin entropy and the uncertainty algorithm. In Sec.~\ref{secNH} we discuss the nonhyperbolic cases, showing the effect of KAM islands on the unpredictability of exit basins of the H\'{e}non-Heiles system and the standard map with two symmetrical exits. In Sec.~\ref{secH} we discuss the hyperbolic case using the four-hill system as an example. Finally, in Sec.~\ref{sec: Conclusions}, we present the main conclusions. 

\section{Unpredictability measures: basin entropy and uncertainty algorithm} \label{secUM}

The method to compute the basin entropy is as follows. We subdivide the exit basins into a grid composed of $N$ square boxes of linear size $\varepsilon$. Each box is filled with $n_t$ trajectories (25 in our case), to each of them we associate a natural number depending on the destination of the particle. In order to plot the exit basins we associate a color to each natural number. Using this convention, the entropy of a certain box $i$ is given by
\begin{equation} \label{eq:Sb1}
S_i = \sum_{j=1}^{c_i}\frac{n_{i,j}}{n_t}\log{\left(\frac{n_t}{n_{i,j}}\right)},
\end{equation}
where $c_i$ is the number of different colors in the box $i$ and $n_{i,j}$ is the number of points with color $j$ in the box $i$. The quotient $n_{i,j}/n_t$ is the probability of the color $j$. The base of the logarithm is $e$.
\\\indent We calculate the entropy of $N$ square boxes, following a Monte Carlo method, and we compute the total entropy of the exit basin
\begin{equation} \label{eq:Sb2}
S = \sum_{i=1}^{N}S_i=\sum_{i=1}^{N}\sum_{j=1}^{c_i}\frac{n_{i,j}}{n_t}\log{\left(\frac{n_t}{n_{i,j}}\right)}.
\end{equation}
\indent Finally, the basin entropy is defined as the entropy relative to the number of boxes used in the random sampling
\begin{equation} \label{eq:Sb3}
S_b = \frac{S}{N}.
\end{equation}
\\\indent The previous description of the basin entropy gives us an understanding about the computational methods used to obtain it. However, in order to get a better understanding about the factors that affect the basin entropy, we can look at it from another perspective. Let's consider that the colors inside the boxes are equiprobable, so $n_i/n_t=1/c_i$ in any box. Hence, the total entropy reads:
	\begin{equation} \label{eq:basinen}
	S = \sum_{i=1}^{N}\log{c_i}.
	\end{equation}
		\\\indent
	Only the $N_k$ boxes that lie in the boundary between two or more basins contribute to the total entropy, being $k\in[1,k_{max}]$ the label for different boundaries. Hence, we can write the total entropy 
	\begin{equation} \label{eq:basinen2}
S=\sum_{k=1}^{k_{max}}N_k\log{c_k},
	\end{equation}
	and the basin entropy
		\begin{equation} \label{eq:basinen3}
		S_b=\sum_{k=1}^{k_{max}}\frac{N_k}{N}\log{c_k}.
		\end{equation}
		\\\indent The number of boxes of linear size $\varepsilon$ required to cover the boundary $k$ grows as $N_k=n_k\varepsilon^{\alpha_k-D}$\cite{McDonald85}, being $\alpha_k$ the uncertainty exponent, $D$ the dimension of the phase space, and $n_k>0$ a constant. On the other hand, the number of boxes required to cover all the phase space grows like $N=\tilde{n}\varepsilon^{-D}$, where $\tilde{n}>0$ is a constant. Using these formulas for $N$ and $N_k$ in Eq. (\ref{eq:basinen3}) we obtain
\begin{equation} \label{eq:Sb}
S_b = \sum_{k=1}^{k_{max}}\frac{n_k}{\tilde{n}}\varepsilon^{\alpha_k}\log{c_k}.
\end{equation}
\\\indent
Although we never use this equation to compute the basin entropy, we can get from it a qualitative information of the different ingredients that affect the basin entropy, that is,  the size of the boundaries ($n_k/\tilde{n}$), the uncertainty dimension ($\varepsilon^{\alpha_k}$) and the number of colors in the basins ($\log{c_k}$).\\\indent
The way to perform the calculation of the fractal dimension is the following. We obtain the exit for a certain initial condition $(x_0,y_0)$, and also the exit for the weakly perturbed initial conditions $(x_0+\delta,y_0)$, $(x_0-\delta,y_0)$, $(x_0,y_0+\delta)$ and $(x_0,y_0-\delta)$. If all of them coincide we will say that the initial condition is certain. On the other hand, if they do not coincide we will label the initial condition as uncertain. We repeat this procedure for many initial conditions and many values of the perturbation $\delta$, and we calculate the fraction of uncertain initial conditions, that obeys the power law:
\begin{equation} \label{fractionuncertain}
f(\delta) \sim \delta^{\alpha},
\end{equation}
where $\alpha=D-d$ is the uncertainty exponent, beeing $D$ the dimension of the phase space and $d$ the fractal dimension. 
\\\indent Taking logarithms in the above equation we obtain
\begin{equation} \label{logfrac}
\log{f(\delta)} = (D-d)\log{\delta}+c,
\end{equation}
where $c$ is a constant. Using this equation we can obtain the fractal dimension $d$ computationally from the slope of the line that must yield a plot of $\log{f(\delta)}$ versus $\log{\delta}$.
\\\indent
In all the simulations of this manuscript related to the fractal dimension calculation, we have taken $250000$ initial conditions in order to obtain the fraction of uncertain initial conditions for each $\delta$. On the other hand, we have taken 21 values of $\delta$ from $10^{-9}$ to $10^{-5}$.

\section{Nonhyperbolic case} \label{secNH}

\begin{figure}
	\centering
	\includegraphics[width=0.47\textwidth,clip]{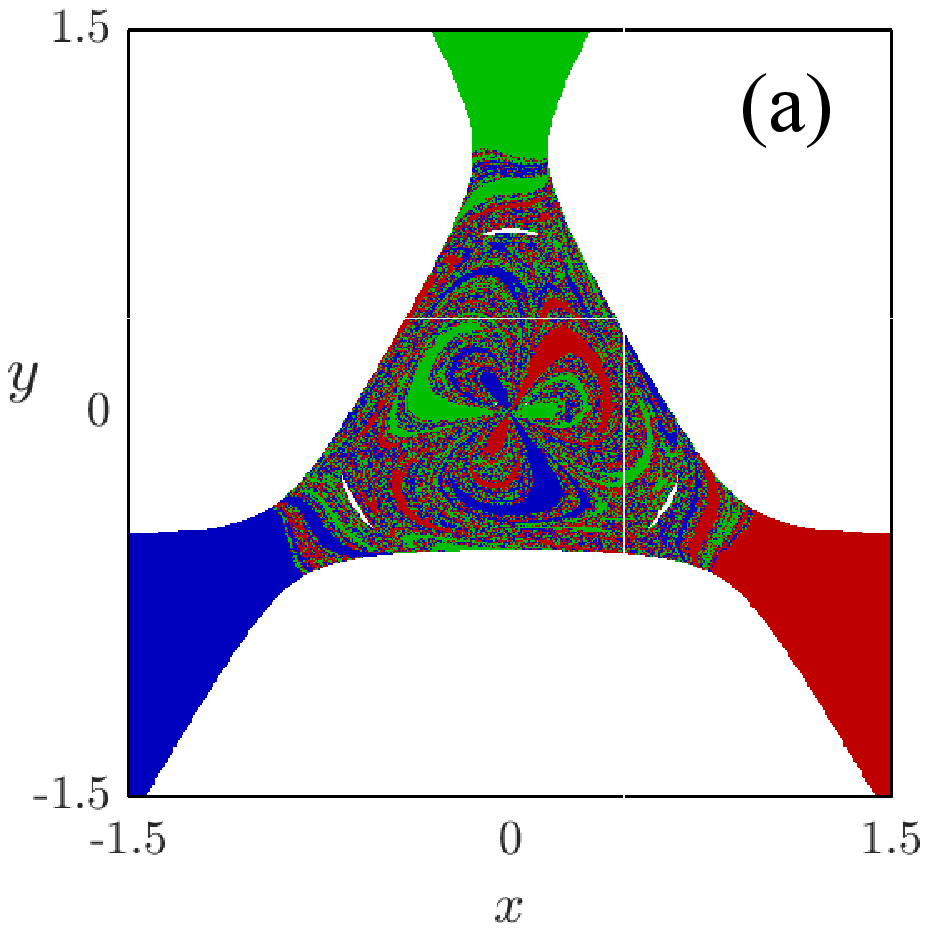}
	\includegraphics[width=0.47\textwidth,clip]{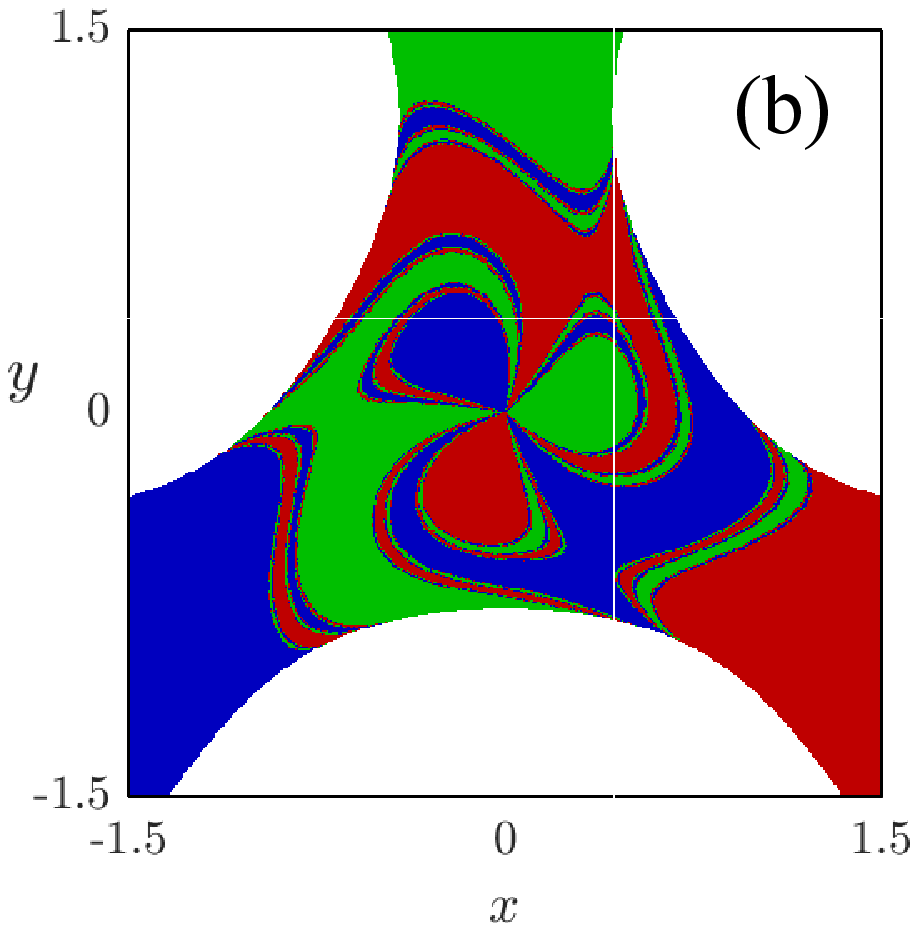}
	\caption{Exit basins in the physical space for the H\'{e}non-Heiles system. The energies are (a) $E=0.20$ and (b) $E=0.45$. The colors red, green and blue refer to initial conditions leading to the three exits shown in the figure. The white color in panel (a) corresponds to the bounded orbits that never escape, and make up the KAM islands. Since there are no KAM islands in panel (b), hence the system is hyperbolic for this value of the energy}
	\label{fig:basins}
\end{figure}

Perhaps the most significant model discussed in this manuscript is the H\'{e}non-Heiles system \cite{HH64}. This system arose in the context of celestial mechanics and is given by the Hamiltonian 
\begin{equation} \label{eq:HH_Hamiltonian}
{\cal{H}}=\frac{1}{2}(\dot{x}^2+\dot{y}^2)+\frac{1}{2}(x^2+y^2)+x^2y-\frac{1}{3}y^3.
\end{equation}
\indent The system becomes a paradigmatic example of chaotic scattering if the energy is higher than the threshold value $E_e =1/6$. Over this value of the energy the isopotential curves are open and hence, the particles can escape from the scattering region through one of the three exits of the potential well. To intuitively visualize the system, we show in Fig.~\ref{fig:basins} the exit basins in the physical space $(x,y)$, following the tangential shooting method \cite{Aguirre01}, for different values of the energy. The exit basins of the H\'{e}non-Heiles system have been studied in many works (e.g., Refs. \cite{Aguirre01,Aguirre03,Blesa12,Barrio08}). For the energy value used in panel (a), the system is nonhyperbolic and has KAM islands (see white regions) mixed with the exit basins. We cannot observe KAM islands in the basin of panel (b), because  the basin has been computed for an energy value in which the system is hyperbolic.

\begin{figure}[htp]
	\centering
	\includegraphics[width=0.46\textwidth,clip]{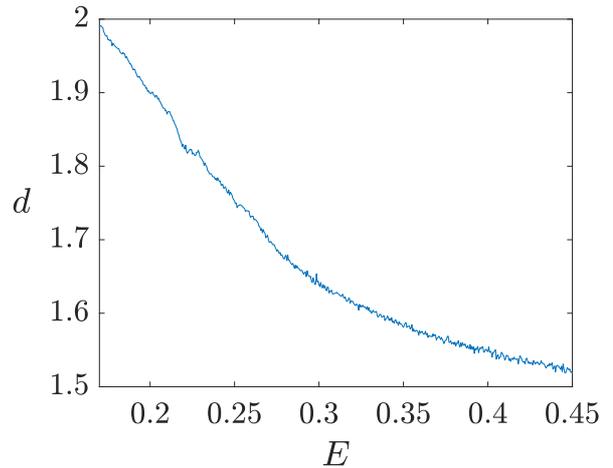}
	\caption{Variations of the fractal dimension of the basin boundaries of the H\'{e}non-Heiles system with the energy. For each energy and for each value of $\delta$, $250000$ initial conditions have been launched in order to compute the fractal dimension }
	\label{fig:dHH}
\end{figure}

\indent First, we have computed the fractal dimension $d$. The evolution of $d$ with increasing energy is shown in Fig.~\ref{fig:dHH}. In the figure we can see that $d$ decreases monotonously with $E$. The result is intuitive, since increasing the energy also increases the size of the exits and reduces the escape times. Consequently in the exit basins we can observe the decrease of the width and fractality of the basin boundaries (see, for example, Fig.~\ref{fig:basins}). 

\begin{figure}[htp]
	\centering		
	\includegraphics[width=0.48\textwidth,clip]{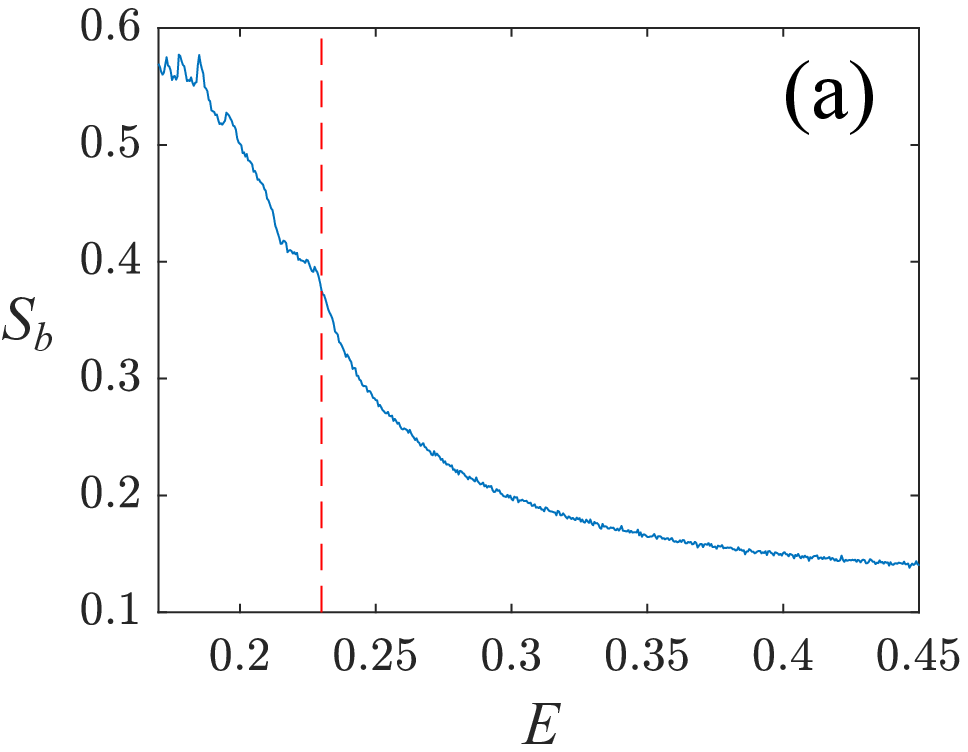}
	\includegraphics[width=0.48\textwidth,clip]{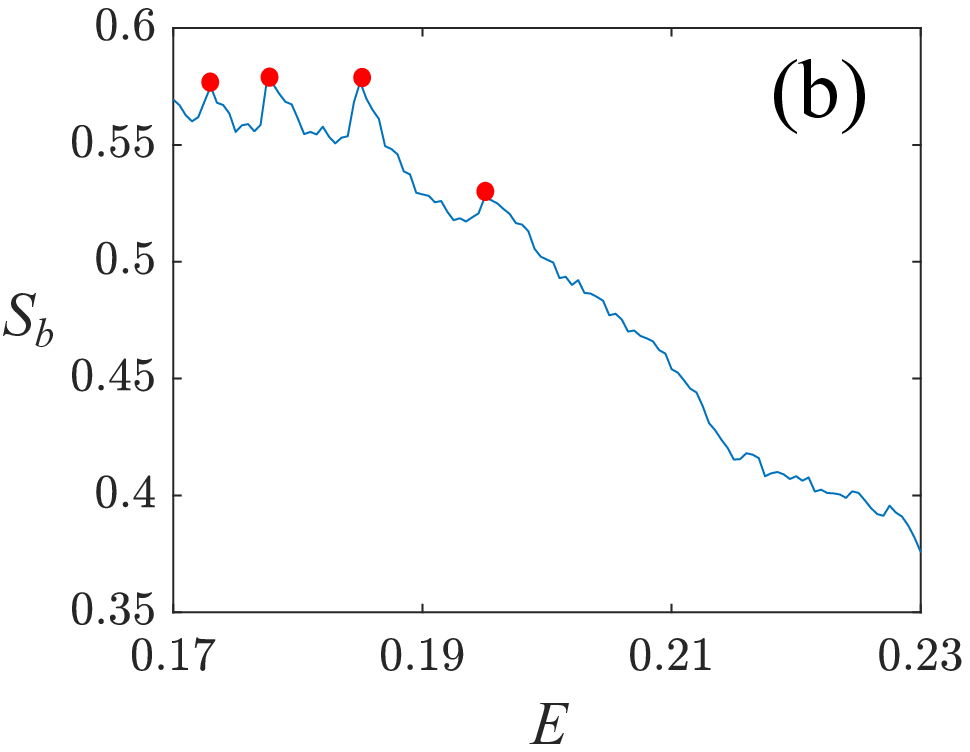}
	\caption{(a) Basin entropy evolution of the exit basins of the H\'{e}non-Heiles system with increasing energy. Two different regions can be observed in the figure: fluctuations ($E\in[0.17,0.23]$) and, after a slight jump in the basin entropy, a monotonous decrease ($E>0.23$). The red dashed line is located at $ E=0.2309 $ and separates both regions. (b) Zoom-in of the nonhyperbolic region of the (a) panel, showing four relative maxima of the basin entropy (see red circles), that appear for energies $E=0.1735$, $0.1775$, $0.1850$ and $0.1950$ }
	\label{fig:HH_Sb1}
\end{figure}

\begin{figure}[htp]
	\centering		
	\includegraphics[width=0.47\textwidth,clip]{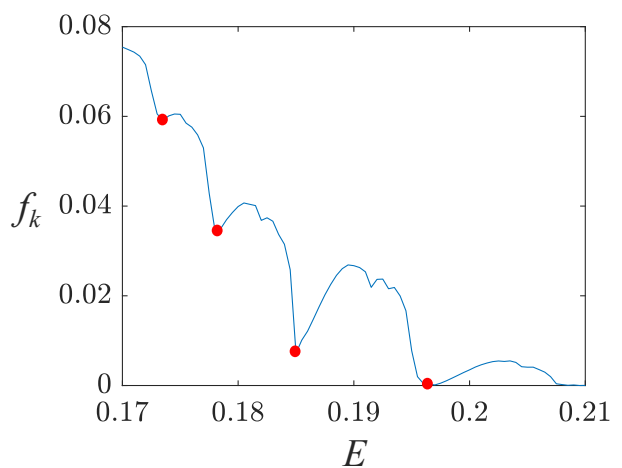}
	\caption{Fraction of the area of the exit basins occupied for the KAM islands in function of the energy of the H\'{e}non-Heiles system. The four relative minima (see red circles) occur when the energy is $E=0.1735$, $0.1775$, $0.1850$ and $0.1950$, respectively. These values correspond to the values of the energy that generate relative maxima in the basin entropy }
	\label{fig:HH_fkam}
\end{figure}

\begin{figure*}[htp]
	\centering		
	\includegraphics[clip=true,width=0.8\textwidth,clip]{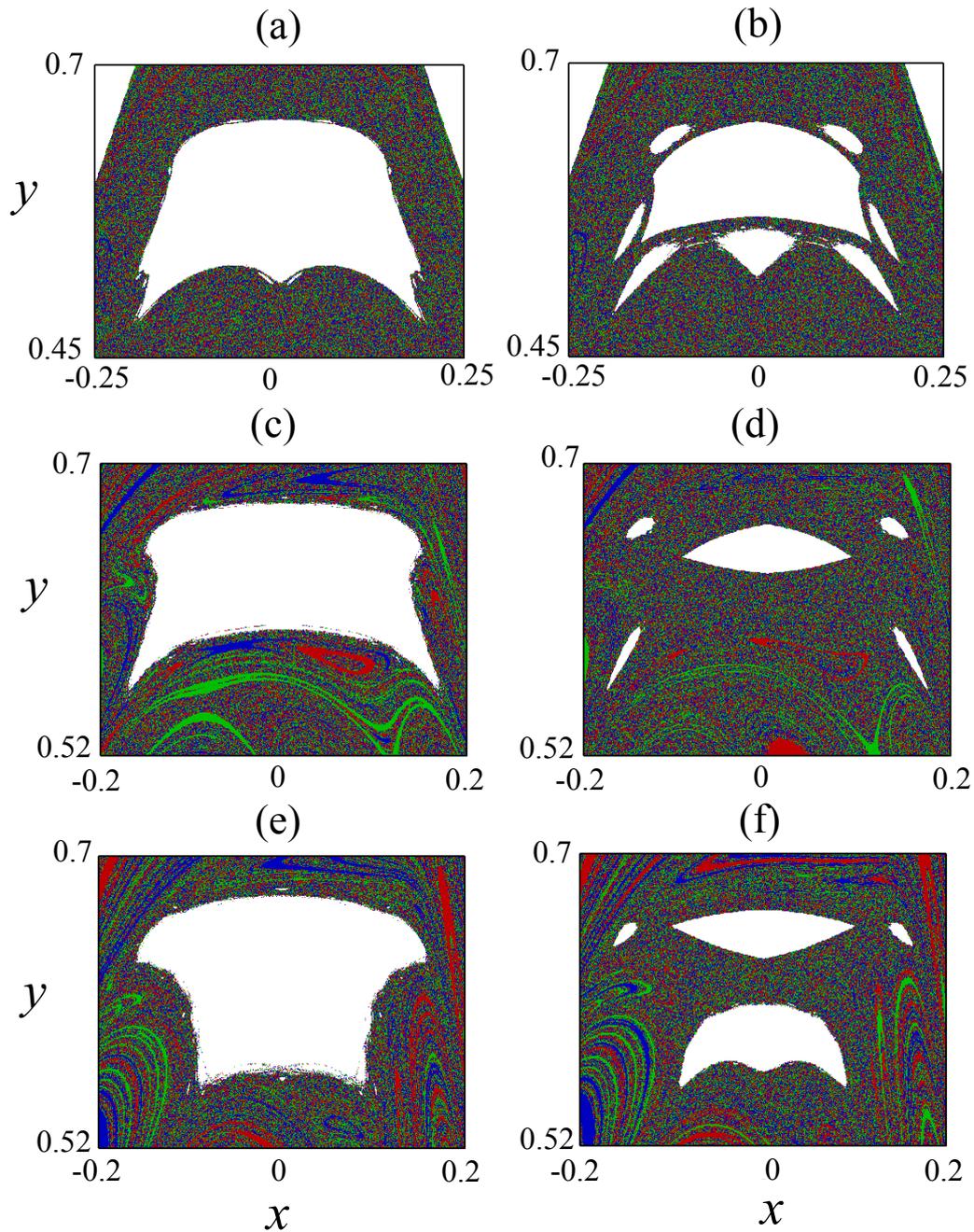}
	\caption{Zoom-in of the exit basins, showing the KAM islands in the physical space for energies (a) $0.1770$, (b) $0.1775$, (c) $0.1845$, (d) $0.1850$, (e) $0.1945$ and (f) $0.1950$. The white regions are the KAM islands and the other colors refer to the initial conditions leading to escaping trajectories. We can observe big changes in both the size and geometry of the KAM islands when we modify slightly the energy}
	\label{fig:HH_KAM_zoom}
\end{figure*}

\indent The KAM islands in the exit basins constitute regions of high predictability and exhibit a smooth boundary with the other three basins. For this reason the size of the KAM islands does not have deep implications in the fractal dimension. In fact, the evolution of the fractal dimension in Fig.~\ref{fig:dHH} is the same if we do not consider the KAM islands as a different destination of the dynamical system. Since the KAM islands are not mixed with the three exit basins in a complex manner, a big area of the exit basins occupied by the KAM islands leads to a higher predictability of the system. As we mentioned in the introduction, the basin entropy depends on the number of destinations in the exit basins, the boundary size, and its fractality. Therefore the existence of KAM islands, although it does not affect the fractality of the boundaries, does affect the other two ingredients. 
\\\indent In order to compute the exit basins, we have used a $1000 \times 1000$ grid filled with initial conditions in the region $\Omega \in[-1,5,1.5]\times[-1,5,1.5]$. In all our simulations we have used a very long maximum time of integration $t=100000$ red using a fourth-order Runge-Kutta method, in order to ensure that the particles that have not escaped will not escape. We have computed $400$ exit basins for different energies in the range $E\in[0.17,0.45]$. For each exit basin we have obtained the basin entropy after launching $100000$ boxes in the region $\Omega$, following a Monte Carlo method. The result is shown in Fig.~\ref{fig:HH_Sb1}. We can clearly observe two different regions: fluctuations in the basin entropy ($E\in[0.17,0.23]$) and a monotonous decrease ($E>0.23$). The first region coincides with the nonhyperbolic regime, while the second corresponds to the hyperbolic regime. Previous research reported that the KAM islands disappear on the $y$-axis around $E\approx 0.2113$ \cite{Barrio08}. Our numerical simulations concerning the size of the KAM islands support the result shown by the basin entropy, detecting the disappearance of the KAM islands on the physical space $(x,y)$ for $E\approx 0.2309$.
\\\indent Because the exit basins of the H\'{e}non-Heiles system are Wada \cite{Aguirre01,Kennedy}, there is only one boundary between the exit basins. When the regime is nonhyperbolic, there is a second boundary that separates the KAM islands from the other three basins. Hence, following Eq.~(\ref{eq:Sb}), the basin entropy is given by
\begin{equation} \label{eq:SbHH}
S_b = \frac{n_1}{\tilde{n}}\varepsilon^{\alpha_1}\log{3}+\frac{n_2}{\tilde{n}}\varepsilon^{\alpha_2}\log{4},
\end{equation}
where the first term refers to the boundary of the exit basins (with only 3 possible destinations) and the second term with the boundary between the exit basins and the KAM islands (with 4 possible destinations).
\\\indent Since the fractal dimension decreases monotonously with an increasing value of the energy, the fluctuations in the basin entropy must be related to the term $n_k/\tilde{n}$, that is, to the size of the boundaries. Because the KAM islands in the exit basins are regions of high predictability, we can guess that the larger these are, the lower the basin entropy (as long as the other factors remain constant). The second term of Eq.~(\ref{eq:SbHH}) will increase if the size of the KAM islands increases. However, this term will have little weight in the final value of the basin entropy, since the boundary of the KAM islands is much lower than the boundary of the exit basins. Therefore, the main effect of an increase in the area occupied by the KAM islands is the decrease in the size of the boundaries between the exit basins, which implies a decrease in the basin entropy. To verify the above arguments, we have calculated the fraction of the exit basins occupied by KAM islands in terms of the energy. The result is shown in Fig.~\ref{fig:HH_fkam}. The maximum value of the energy shown in the figure is $E=0.21$, since the fraction occupied by the KAM islands is very small for higher values of the energy. However, as we mentioned before, the disappearance of the KAM islands occurs in $E\approx 0.2309$. 
\\\indent The relative maxima in the basin entropy that were observed in Fig.~\ref{fig:HH_Sb1} correspond to the abrupt decrease in the size of the KAM islands. To illustrate these changes in the KAM islands we represent a zoom-in of the exit basins in Fig.~\ref{fig:HH_KAM_zoom}, showing the KAM islands for very close values of the energy. These values correspond to the second, third and fourth relative maxima of Fig.~\ref{fig:HH_Sb1}(b). A metamorphosis can be observed in Fig.~\ref{fig:HH_KAM_zoom} from the left panel to the right by a small variation ($0.005$) in the energy.

%

\begin{figure*}[htp]
	\centering		
	\includegraphics[trim=0cm 0cm 0cm 0 cm,clip=true,width=0.4\textwidth,clip]{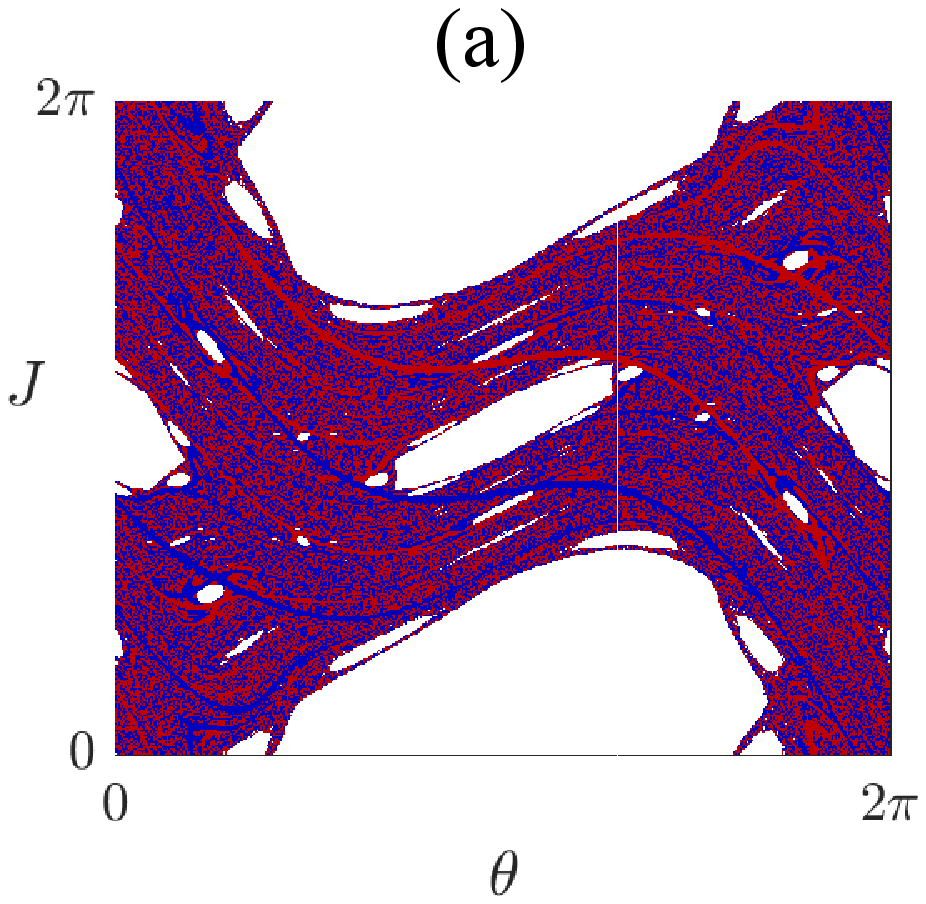}
	\includegraphics[trim=0cm 0cm 0cm 0 cm,clip=true,width=0.4\textwidth,clip]{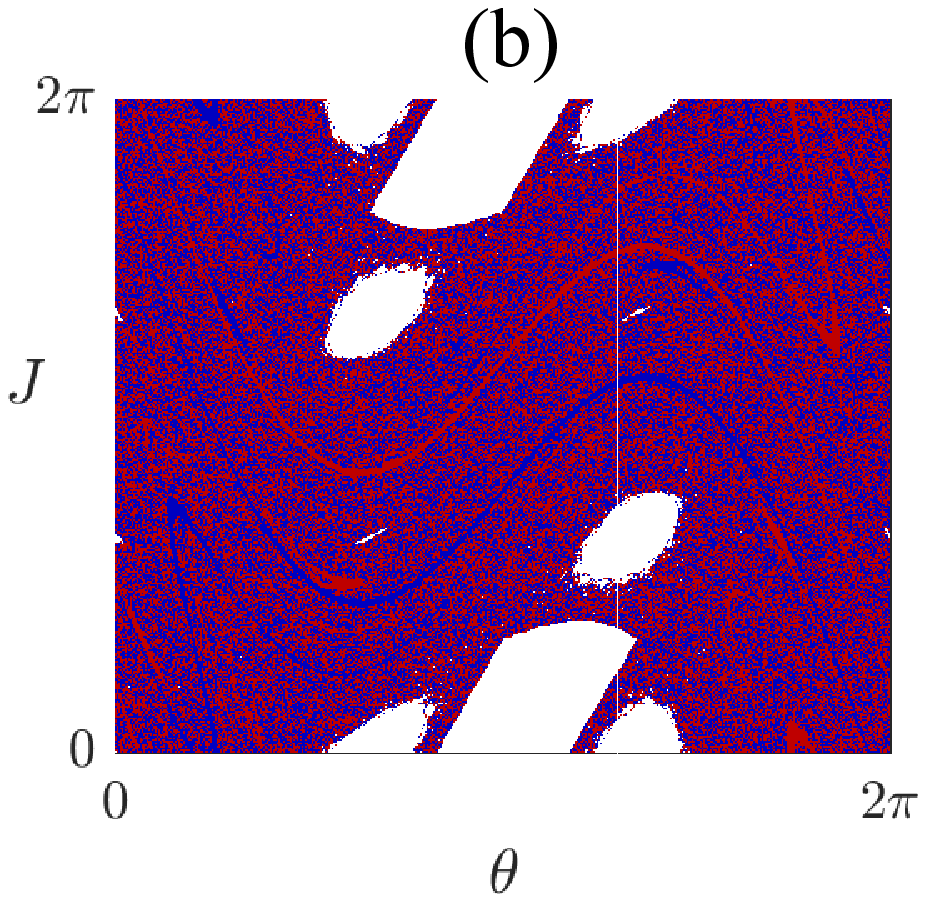}	
	\includegraphics[trim=0cm 0cm 0cm 0 cm,clip=true,width=0.4\textwidth,clip]{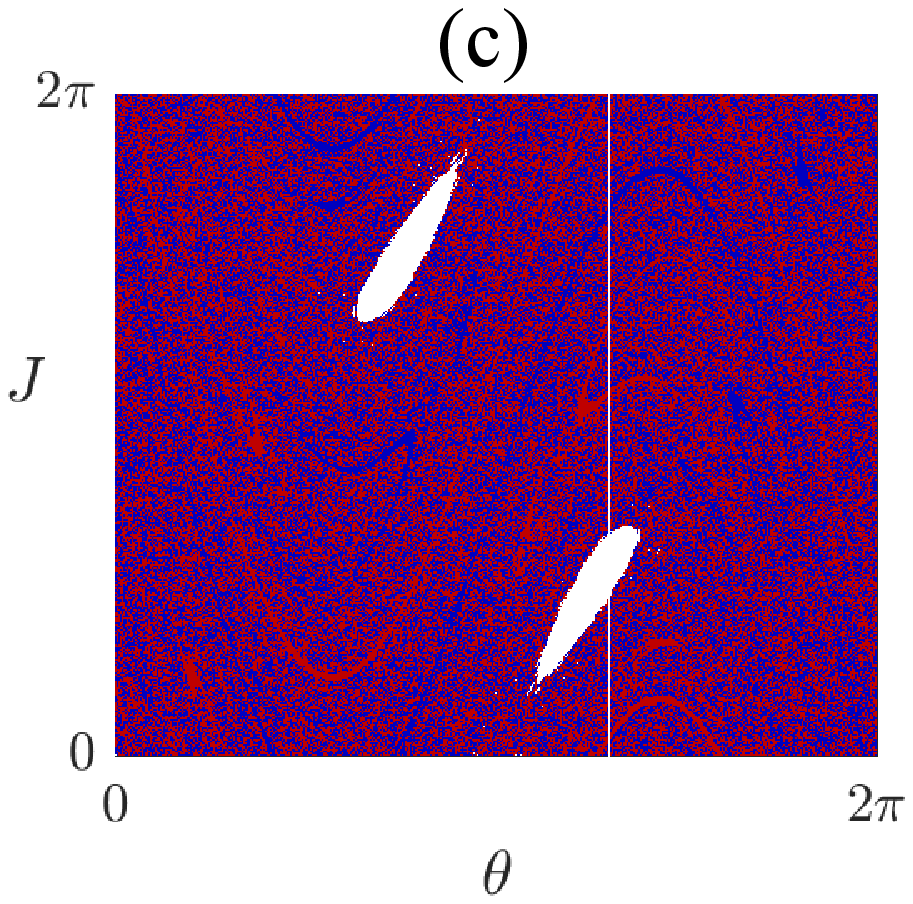}
	\includegraphics[trim=0cm 0cm 0cm 0 cm,clip=true,width=0.4\textwidth,clip]{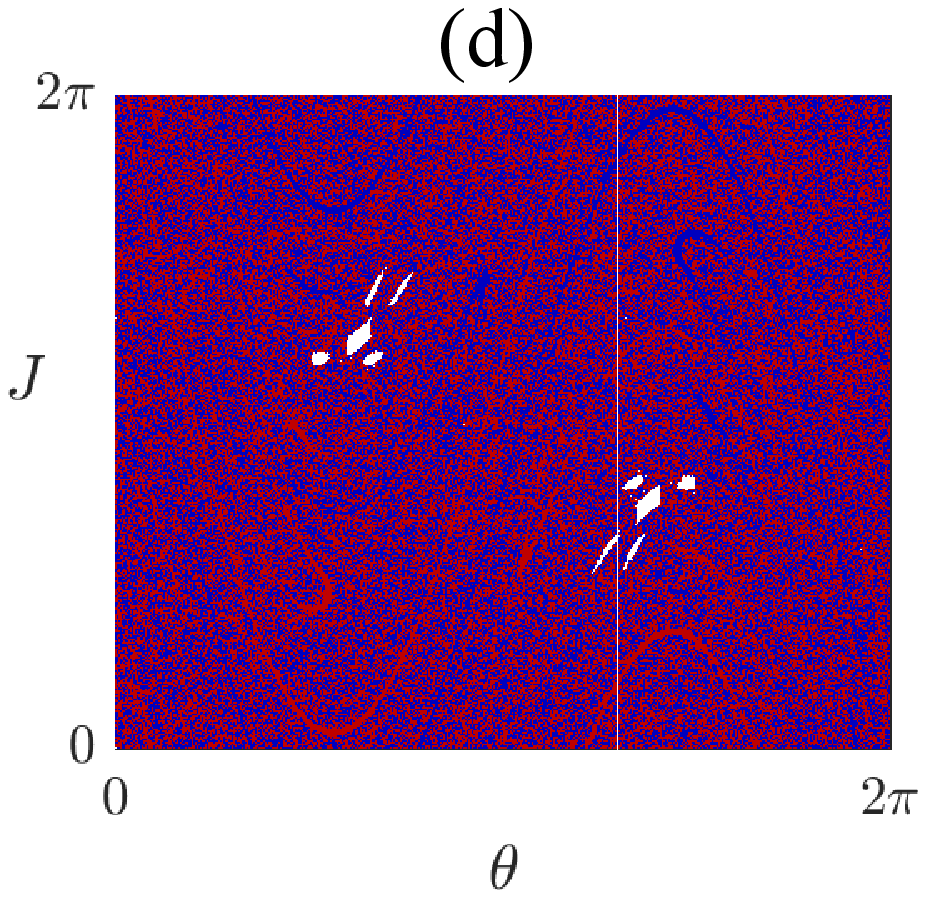}
	\caption{Exit basins of the standard map with two symmetrical exits of width $\omega=0.01$, for different values of the parameter (a) $K=1.2$, (b) $K=2.45$, (c) $K=4.5$ and (d) $K=5.1$. The color code is as follows: red and blue correspond to the initial conditions that lead to the left exit and the right exit, respectively. White colors refer to bounded orbits, and hence white regions are the KAM islands. Variations in the size and geometry of the KAM islands can be observed }
	\label{fig:SM_KAM}
\end{figure*}

\begin{figure}[htp]
	\centering		
	\includegraphics[width=0.47\textwidth,clip]{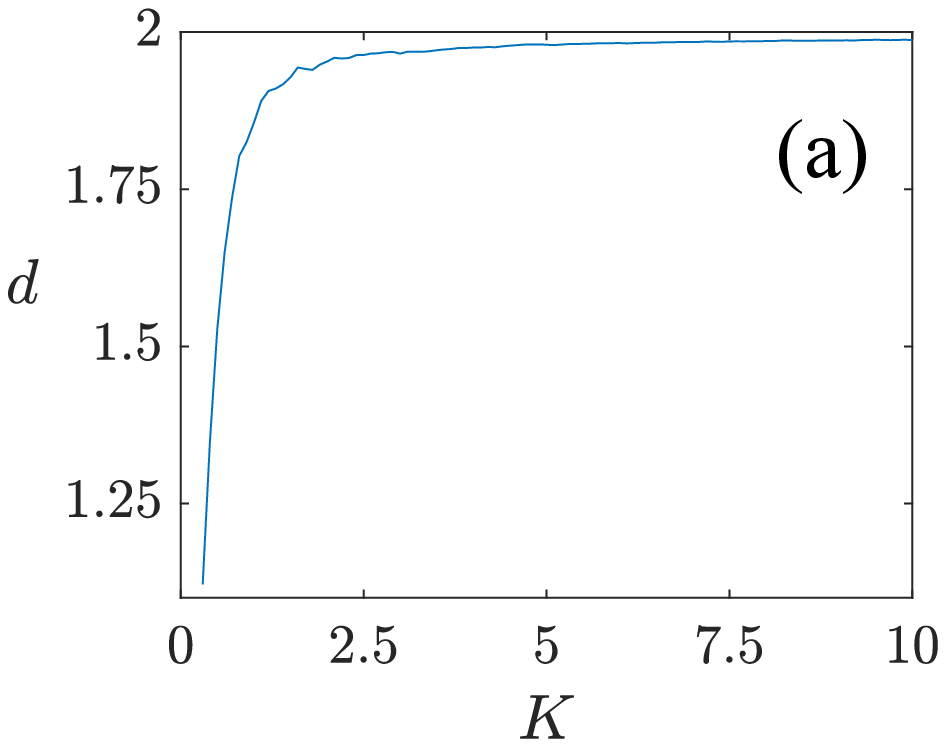}
	\includegraphics[width=0.47\textwidth,clip]{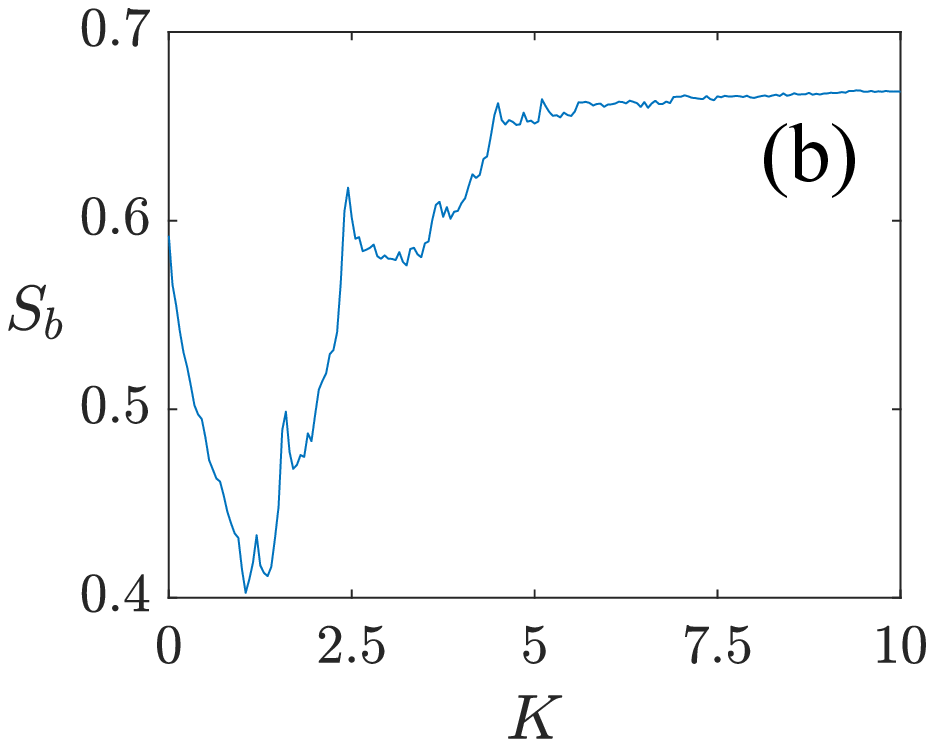}
	\caption{Fractal dimension and basin entropy evolution of the exit basins of the standard map with two symmetrical exits of width $\omega=0.01$. (a) The fractal dimension increases monotonously. (b) The basin entropy exhibit fluctuations due to the effect of KAM islands }
	\label{fig:Ch_Sb}
\end{figure}

\indent In the hyperbolic regime, due to the absence of KAM islands, there exists only one boundary in the exit basins and hence the basin entropy is given by
\begin{equation} \label{eq:SbHH2}
S_b = \frac{n}{\tilde{n}}\varepsilon^{\alpha}\log{3}.
\end{equation}
\indent As the energy increases, the escape times of the trajectories are reduced, so that the boundary becomes thinner and the term $n/\tilde{n}$ decreases monotonously. Since the fractal dimension also decreases monotonously with increasing energy, we do not appreciate fluctuations in the basin entropy in the hyperbolic regime. 
\\\indent This result allows us to use the basin entropy as a tool to discern whether the regime is hyperbolic or nonhyperbolic. Moreover, in systems in which the regime changes in some value of the energy (or another parameter), we can detect this change by looking for a jump in the basin entropy. This result is quite important, because in previous research the method used to find the transition between regimes was based on the decay law of the survival probability \cite{Blesa14,Bernal13}. The method based on the basin entropy requires less computational effort and is more accurate. 
\\\indent Finally, the minimum value reached by the basin entropy for $E=0.45$ is $S_b\approx0.13$, as can be observed in Fig.~\ref{fig:HH_Sb1}. If we increase even more the energy, the boundary will continue becoming thinner. For this reason we can expect that for very high values of the energy
\begin{equation} \label{eq:SbHH3}
\lim_{E\to\infty}\frac{n}{\tilde{n}}=0 \implies \lim_{E\to\infty}S_b=0.
\end{equation}
\indent In order to show similar conclusions in a different system, we have computed the basin entropy and the fractal dimension of the exit basins of the standard map \cite{Chirikov} with two symmetrical exits. The equation of the standard map is given by

\begin{equation} \label{eq:standard}
\begin{aligned}
\theta_{n+1} & = \theta_n + J_{n+1} \quad \mbox{mod} \kern 0.2pc 2\pi,\\
J_{n+1} & = J_n + K\sin{\theta_n},			
\end{aligned}
\end{equation}
where $K>0$ is a constant. 	
\\\indent The system is a closed Hamiltonian map. However, as explained in \cite{opening}, it is possible to open the system by introducing exits. These exits represent some kind of interaction with the outside, and allow us to construct the exit basins to study the underlying dynamics of the system. The procedure is as follows. We define two regions $E_1\equiv[\theta_1,\theta_2]\times[0,2\pi]$ and $E_2\equiv[\theta_3,\theta_4]\times[0,2\pi]$. Arbitrarily, we place the center of the regions in $\theta=0.2\pi$ ($E_1$) and in $\theta=1.8\pi$ (in order to be located at the same distance of $\theta = 0$). The width of each region is $\theta_2-\theta_1=\theta_4-\theta_3=2\pi w$, where $w\leq 0.2$ is the parameter that we can use to modify the size of the regions. Following this method the left region is placed in $[0.2\pi-w\pi,0.2\pi+w\pi]\times[0,2\pi]$ and the right region in $[1.8\pi-w\pi,1.8\pi+w\pi]\times[0,2\pi]$. If after one or more iterations the orbit falls in $E_1$ or $E_2$, we say that the orbit has escaped. Following this procedure for several initial conditions in the $(\theta,J)$ plane, and assigning blue color to the exit 1 and red color to the exit 2, we can construct the exit basins. The arbitrary choice of the shape and width of the exits does not affect to the size and geometry of the KAM islands.  If we define exits with different size or geometry, only the fractal dimension of the basin boundaries would change, but the results regarding the basin entropy would be qualitatively identical.
\\\indent To carry out the computation of the exit basins, we have used a $1000\times 1000$ grid filled with initial conditions in the region $[0,2\pi]\times[0,2\pi]$. The parameter that we have varied is $K$. As an example, we show in Fig.~\ref{fig:SM_KAM} the exit basins for different values of $K$.
\\\indent Figure \ref{fig:SM_KAM} shows the existence of large KAM islands for low values of $K$. As the value of the parameter increases, the KAM islands evolve, changing their size and geometry. Although the fractal dimension increases monotonously until it stabilizes at the maximum possible value $ d = 2 $ (see Fig.~\ref{fig:Ch_Sb} (a)), the basin entropy follows a totally different trend, as shown in Fig.~\ref{fig:Ch_Sb} (b). The fluctuations in the evolution of the basin entropy stop after the value $ K \approx 7.5 $, because the size of the KAM islands is extremely small and then the dominating term is the fractal dimension.

\section{Hyperbolic case}  \label{secH}

\begin{figure}[htp]
	\centering
	\includegraphics[width=0.42\textwidth,clip]{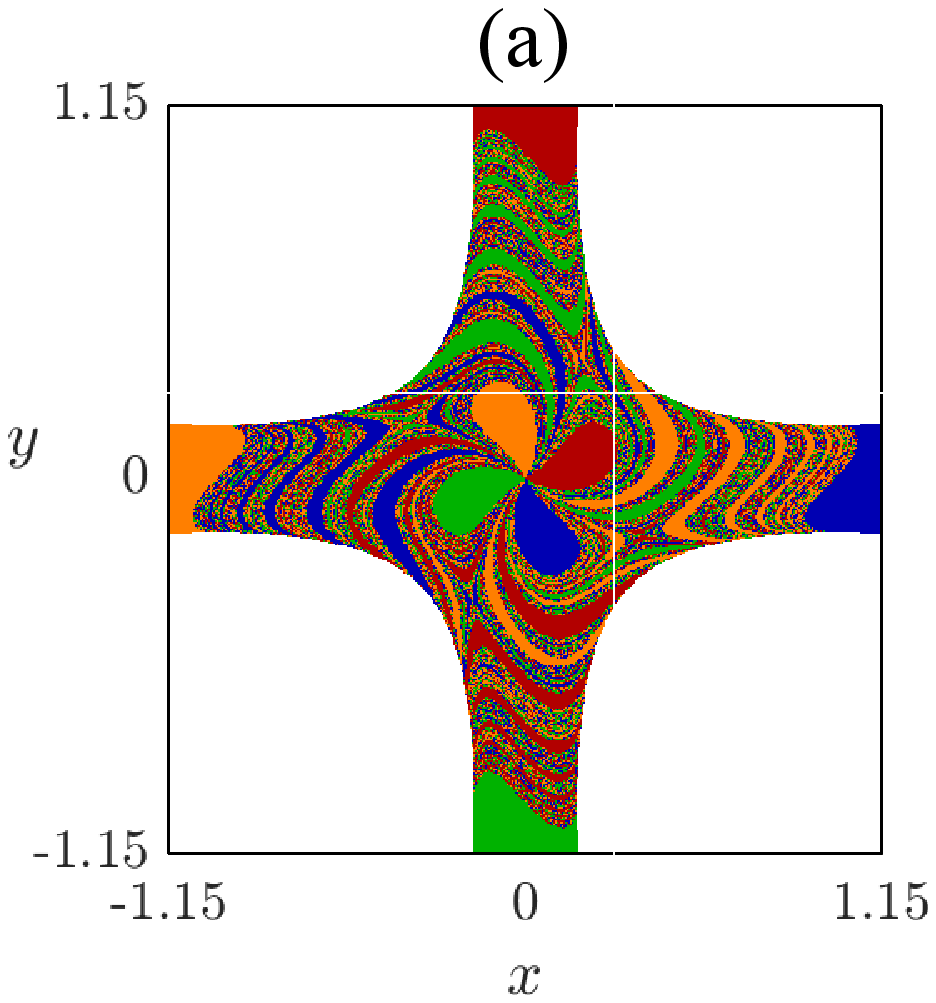}
	\includegraphics[width=0.42\textwidth,clip]{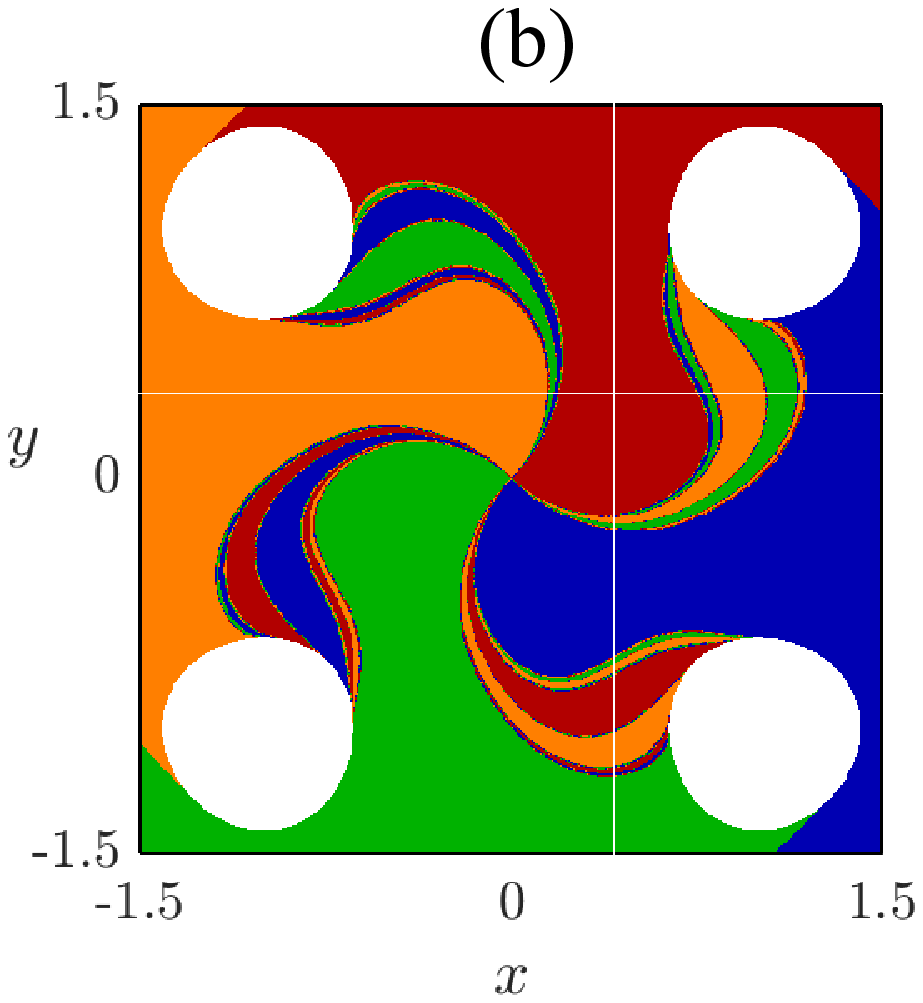}
	\caption{ Exit basins in the physical space of the four-hill system. The energies are (a) $E=0.01$ and (b) $E=0.1$. The different colors refer to initial conditions leading to the four different exits of the Hamiltonian. We can clearly observe that the dynamics is much more unpredictable for $E=0.01$ than for $E=0.1$ }
	\label{fig:4Hbasins}
\end{figure}

Our numerical example of a hyperbolic system is the four-hill system \cite{Bleher,Zotos}, given by
\begin{equation} \label{eq:4H_Hamiltonian}
{\cal{H}}=\frac{1}{2}(\dot{x}^2+\dot{y}^2)+x^2y^2e^{-(x^2+y^2)}.
\end{equation}
\indent The potential of the system consists of four hills located at $(x,y)=(\pm 1, \pm 1)$. For any value of the energy the isopotential curves are open and the particles can escape through four symmetrical exits, separated by an angle of $\pi/2$ radians. We have chosen this system because it has two interesting characteristics. First, the system has a maximum value of the energy $E_m=1/e^2\approx0.135$ above which the scattering is nonchaotic \cite{Seoane13}, and hence the basin boundary becomes smooth. On the other hand, in the range $E\in(0,E_m]$ the scattering is always hyperbolic. For this reason there are no KAM islands in the exit basins. In order to visualize the system we plot two exit basins for energies  $E=0.01$ and $E=0.1$ in Fig.~\ref{fig:4Hbasins}.  

\begin{figure}[htp]
	\centering
	\includegraphics[width=0.45\textwidth,clip]{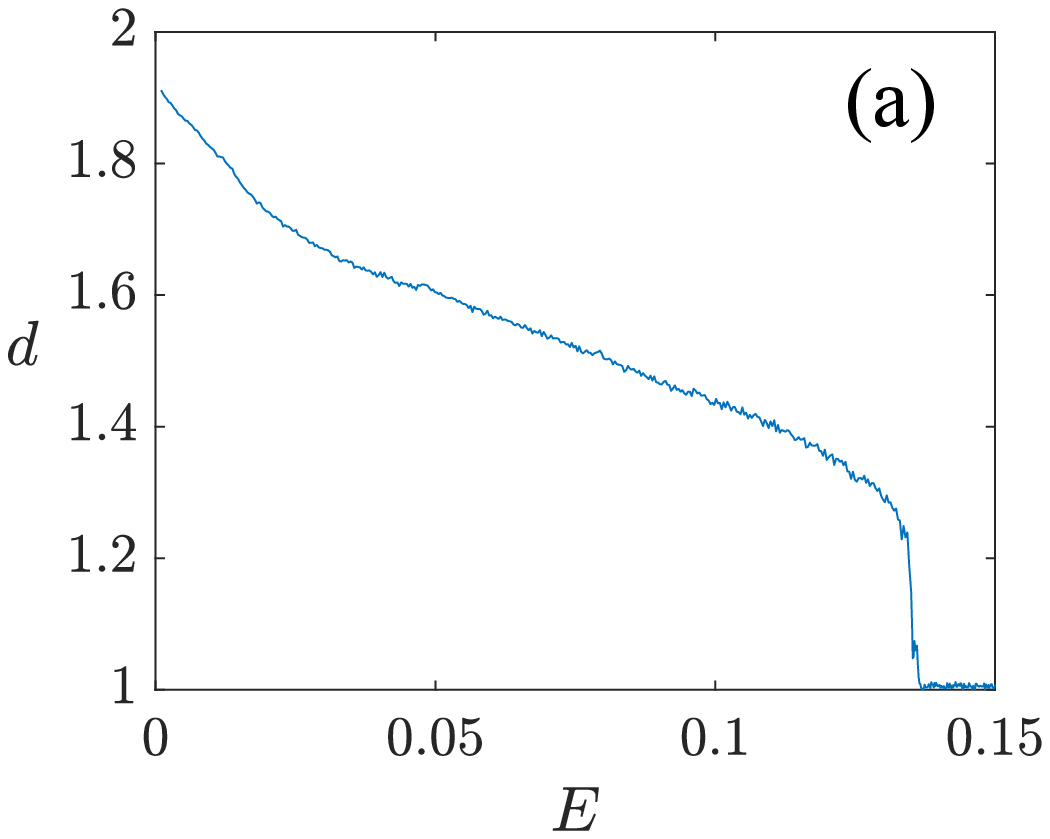}
	\includegraphics[width=0.45\textwidth,clip]{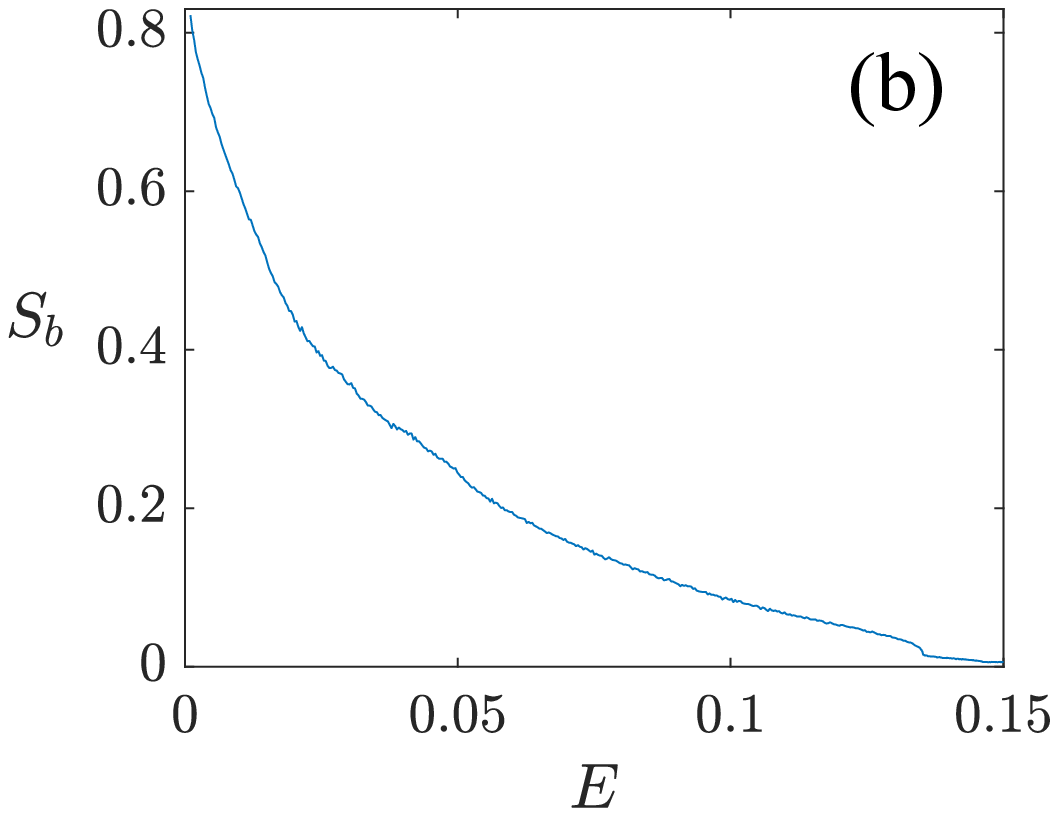}
	\caption{(a) Fractal dimension and (b) basin entropy of the exit basins of the four-hill system for different values of the energy. For each energy and for each value of $\delta$, $250000$ initial conditions have been launched in order to compute the fractal dimension. To compute the basin entropy $500$ basins of resolution $1000\times 1000$ inside the region $\Omega\in[-1.5,1.5]\times[-1.5,1.5]$ have been computed. For each exit basin, the basin entropy has been computed using a random sampling with $N_{in}=100000$ boxes inside the potential  }
	\label{fig:4HdS}
\end{figure}

\begin{figure}[htp]
	\centering
	\includegraphics[width=0.47\textwidth,clip]{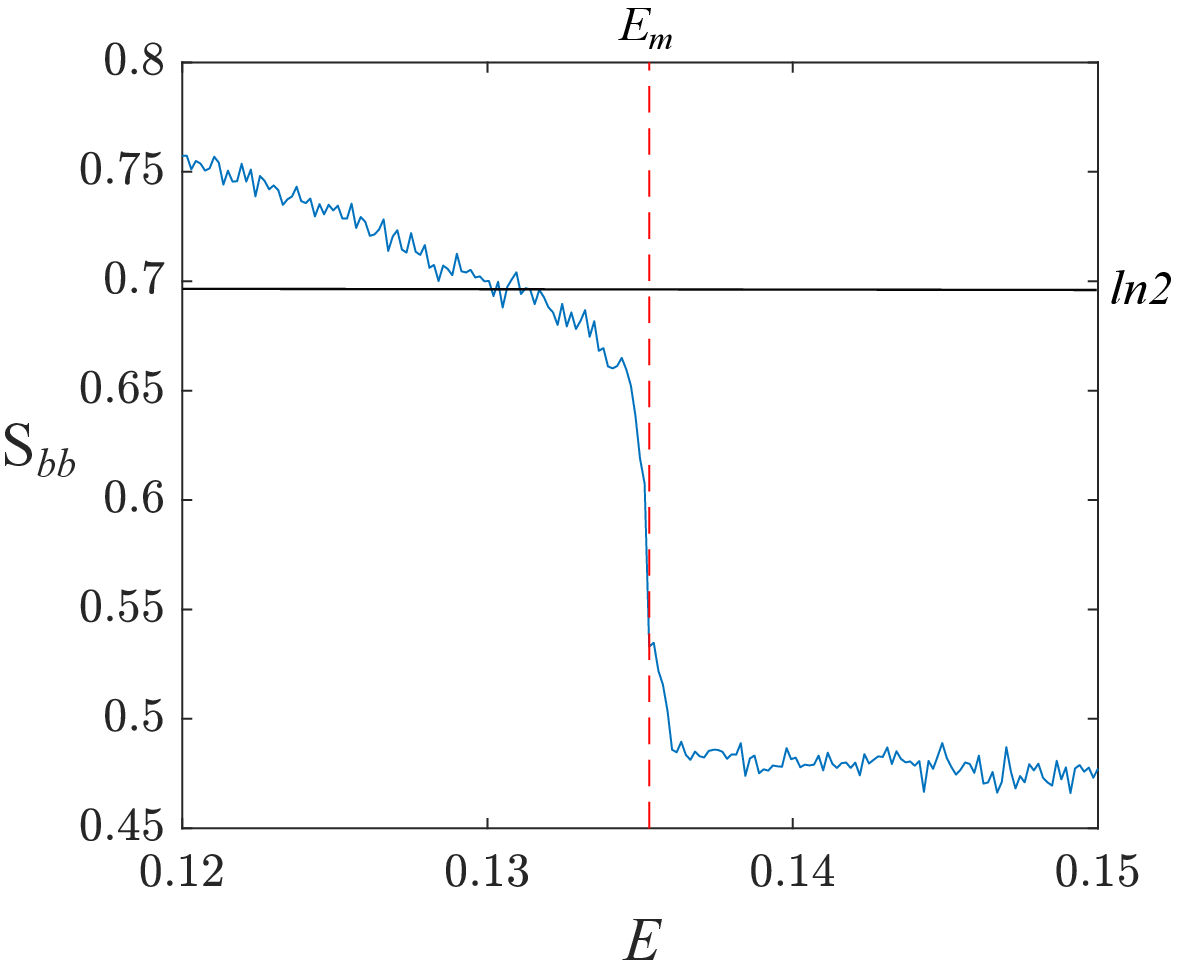}
	\caption{Boundary basin entropy of the exit basins of the four-hill near the critical value $E_m=1/e^2\approx 0.135$. The horizontal black line is located at the value $ S_ {bb} = \log2 $, while the vertical red line is at $E_m $. The result shows that the boundary basin entropy allows us to detect the transition between fractal and smooth boundaries  }
	\label{fig:sbb}
\end{figure}

Because there are no KAM islands in the exit basins of this system, both the basin entropy and the fractal dimension decrease without fluctuations, as shown in  Fig.~\ref{fig:4HdS}. The main qualitative difference between both figures is the abrupt decrease in the fractal dimension near the value $E=E_m$. When this value of the energy is reached the scattering becomes nonchaotic and the fractal dimension falls to the value $d=1$. This metamorphosis between fractal and smooth boundaries is not strongly detected by the basin entropy. This is because the metamorphosis is a change that affects only to the basin boundary, and for high values of the energy the boundary is really thin. Therefore, the fact that the boundary of the exit basins is smooth or fractal does not substantially affects the basin entropy. 
However, if we are interested in detecting changes in the basin boundary we can use the boundary basin entropy, $S_{bb}$ \cite{Daza16}. This quantity is obtained by simply dividing the total entropy between the number of boxes that fall in the boundaries of the exit basins. The boundary basin entropy allows us to determine if a boundary is fractal. This criterion, known as $\log2$ criterion \cite{Daza16}, is a sufficient condition that states that if $S_{bb}>\log2$, then the boundary is fractal. In Fig.~\ref{fig:sbb}, we show the $S_{bb}$ evolution near the metamorphosis of the boundary of the exit basins. Near the critical value $E=E_m$, the boundary basin entropy decreases abruptly below the value $\log2$, as expected. 
\\\indent We have carried out the calculations of the fractal dimension and the basin entropy in the case of the sawtooth map with two symmetrical exits \cite{opening}, which is an hyperbolic discrete system. In the same line as in the the four-hill system or in the hyperbolic regime of the H\'{e}non-Heiles system, no qualitative difference between both magnitudes has been observed.

\section{Conclusions } \label{sec: Conclusions}

In summary, our research reveals that it is not possible to understand the unpredictability in nonhyperbolic open Hamiltonian systems without considering the KAM islands. By modifying parameters with physical meaning in discrete and continuous systems, significant changes in the geometry and size of the KAM islands have been uncovered. These changes lead to fluctuations in the unpredictability of the exit basins that are not detected by the fractal dimension. We expect that these changes may appear in many dynamical systems with mixed phase space. However, the methods used in this manuscript will not be of interest when the KAM islands are small enough to be considered irrelevant in the dynamics of the system, or when the parameters of interest do not influence its size and geometry.
\\\indent We have provided theoretical reasoning for the fluctuations, from the point of view of the basin entropy concept. In short, a bigger area occupied by the KAM islands leads to a higher predictability of the exit basins, since these are not mixed in a complex manner with the chaotic sea. So, KAM islands are a source of periodicity and predictability in the exit basins. 

In absence of KAM islands the unpredictability of the exit basins studied here follows an evolution without fluctuations. For this reason, the basin entropy allows us to detect accurately the transition between the hyperbolic and the nonhyperbolic regime. Moreover, using this procedure we can reduce the computational effort, since we do not need to compute the exponent of the decay law of the survival probability. 

Despite the fact that the basin entropy allows a reliable portrait of the unpredictability in presence of KAM islands, if the number of destinations of the system does not change and the boundaries do not undergo a metamorphosis, the evolution of the fractal dimension and the basin entropy will be qualitatively the same. Moreover, if we are interested in studying a metamorphosis in the boundaries of the exit basins, it is more useful to use the fractal dimension or the boundary basin entropy than the basin entropy. 

We think that this work could help, giving new perspectives and tools, to future research concerning nonhyperbolic dynamics in chaotic scattering problems. 

For further developments, we think that could be interesting to use integrity measures like the Anisometric Local Integrity Measure, to study the dynamical integrity of the KAM island in the presence of perturbations such as noise, forcing or asymmetries in the size of the exit.

\begin{acknowledgements} 
	The work of A. R. N., J. M. S., and M. A. F. S. has been financially supported by the Spanish State Research Agency (AEI) and the European Regional Development Fund (ERDF) under Project No. FIS2016-76883-P.
\end{acknowledgements}

\section*{Conflict of Interest:}
The authors declare that they have no conflict of interest.



\end{document}